\documentclass[english,aps,preprint]{revtex4}
\usepackage[T1]{fontenc}
\usepackage[latin9]{inputenc}
\usepackage{graphicx}

\makeatletter

\providecommand{\tabularnewline}{\\}

\@ifundefined{textcolor}{}
{%
 \definecolor{BLACK}{gray}{0}
 \definecolor{WHITE}{gray}{1}
 \definecolor{RED}{rgb}{1,0,0}
 \definecolor{GREEN}{rgb}{0,1,0}
 \definecolor{BLUE}{rgb}{0,0,1}
 \definecolor{CYAN}{cmyk}{1,0,0,0}
 \definecolor{MAGENTA}{cmyk}{0,1,0,0}
 \definecolor{YELLOW}{cmyk}{0,0,1,0}
}

\usepackage{babel}

\makeatother

\usepackage{babel}
\begin{document}

\title{Search for $HZZ'$ couplings at the LHC}

\author{V. Ar\i{}}

\email{vari@science.ankara.edu.tr}

\affiliation{Ankara University, Department of Physics, 06100, Ankara, Turkey}

\author{O. Çak\i{}r}

\email{ocakir@science.ankara.edu.tr}

\affiliation{Ankara University, Department of Physics, 06100, Ankara, Turkey}

\author{S. Kuday}

\email{sinankuday@aydin.edu.tr}

\affiliation{Istanbul Aydin University, Department of Electrical and Electronics
Engineering, 34295, Istanbul, Turkey}
\begin{abstract}
New physics models predict the possibility of extra neutral gauge
bosons ($Z'$) associated with an extra $U(1)'$ gauge symmetry. We
study the couplings of the Higgs boson to the $Z$ boson and $Z'$
boson predicted by the new physics models. The couplings of the $Z'$
boson to quarks can also be investigated through the $Z'q\bar{q}$
interactions. The accessible ranges of the parameter space have been
searched for processes $pp\rightarrow HZX$ and $pp\rightarrow HHZX$
at the LHC with $\sqrt{s}=14$ TeV.
\end{abstract}

\pacs{12.60.Fr, 14.80.Cp}

\maketitle

\section{Introduction}

The experimental opportunities can be exploited to search for hints
beyond the standard model (BSM) physics after the recent discovery
of the Higgs boson with a mass of 125 GeV by the ATLAS and CMS experiments
\cite{ATLAS2012,CMS2012} at the LHC. Due to its gauge charges the
Higgs boson may interact with the BSM fields. This new interaction
can also modify the couplings between the Higgs and SM fields at tree
level or loop level. Within some extensions of the standard model
(SM) a new neutral gauge boson $Z'$ can be included \cite{London86,Leike99,Langacker09}. 

The phenomenological studies on the $Z'$ boson signatures can be
found in \cite{Barger04,Carena04,Agashe10,Chiang14}. The present
constraints on the $Z'$ mass and couplings from both electron-positron
and hadron-hadron colliders have been presented in \cite{PDG2012}.
The precision measurements at the $Z$ boson resonance lead to the
limit on $Z-Z'$ mixing \cite{Andreev12}. The ATLAS and CMS experiments
at the LHC, running at higher center of mass energy and luminosity,
have updated the Tevatron limits on the $Z'$ boson mass \cite{Aaltonen2009,Abazov11}.
The ATLAS Collaboration searches for a massive resonance decaying
to top quark pairs in the hadronic channels, and excludes the leptophobic
$Z'$ boson with mass smaller than $1.32$ TeV \cite{ATLAS1211}.
The CMS Collaboration excludes leptophobic $Z'$ boson resonance with
a mass of $m_{Z'}<1.3$ TeV for its width $\Gamma_{Z'}=0.012m_{Z'}$
in the search of heavy resonance decaying into $t\bar{t}$ pair with
subsequent leptonic decays \cite{CMS1211}.

In this work, we investigate the $HZZ'$ couplings via the processes
$pp\rightarrow HZX$ and $pp\rightarrow HHZX$ at the LHC with $\sqrt{s}=14$
TeV. The couplings of the $Z'$ boson to quarks can also be investigated
via $Z'q\bar{q}$ interactions. For the signal process $pp\to HZX$,
the $Z'$ boson contributes in the $s$-channel resonance diagrams
through family diagonal neutral current couplings. However, the signal
process $pp\to HHZX$ has contributions from both the $Z'$-boson
resonance and the $Z$-exchange diagrams. We obtain the accessible
ranges of the mass for two different $Z'$ models at the four-fermion
and six-fermion final states resulting from $H\to b\bar{b}$ and $Z\to l^{+}l^{-}$
decays. These modes exist even if the $Z'$ boson decouples from the
leptons.

\section{The Production Cross Section}

The interaction of the Higgs boson ($H$) with the $Z$ boson and
the new $Z'$ boson can be written through the kinetic term of the
scalar field. The relevant interaction term can be written as $L_{HZZ'}=-g_{Z}g_{Z'}z[H]\upsilon HZ_{\mu}Z'^{\mu}$.
The new boson $Z'$ couples to the fermion field ($f$) with the interaction
term $L_{f\bar{f}Z'}=(g_{Z'}/2)\bar{f}\gamma^{\mu}(C_{V}^{f}-C_{A}^{f}\gamma^{5})f$.
Here, the vector ($C_{V}^{f}$) and axial-vector ($C_{A}^{f}$) couplings
are model dependent. We use a linear change of variable $g_{Z'}=\gamma_{Z'}g_{e}/\cos\theta_{W}\sin\theta_{W}$
for the coupling constant. The triple coupling $HZZ'$ depends on
the $U(1)'$ charge. In the model the triple coupling parameter $z[H]=-1$.
The $U(1)'$ couplings to the fermions for two different $Z'$ models
are given in Table \ref{tab:table1}. Since the extensively studied
Drell-Yan process even constrained the possible $Z'l^{+}l^{-}$ couplings,
we consider two candidate models in which the $Z'$ bosons have small
or no couplings to leptons.

\begin{table}
\caption{Vector and axial-vector couplings of $Z'$ boson predicted by leptophobic
(LP) model and $\eta$ (ETA) model. \label{tab:table1} }

\begin{tabular}{|c|c|c|c|c|c|c|c|c|}
\hline 
 & $C_{V}^{u}$ & $C_{A}^{u}$ & $C_{V}^{d}$ & $C_{A}^{d}$ & $C_{V}^{l}$ & $C_{A}^{l}$ & $C_{V}^{\nu}$ & $C_{A}^{\nu}$\tabularnewline
\hline 
\hline 
$Z'$(LP) & $-7/9$ & $1$ & $2/9$ & $0$ & $0$ & $0$ & $0$ & $0$\tabularnewline
\hline 
$Z'$(ETA) & $0$ & $4/6$ & $1/2$ & $1/6$ & $-1/2$ & $1/6$ & $-1/6$ & $-1/6$\tabularnewline
\hline 
\end{tabular}
\end{table}

\begin{figure}
\includegraphics{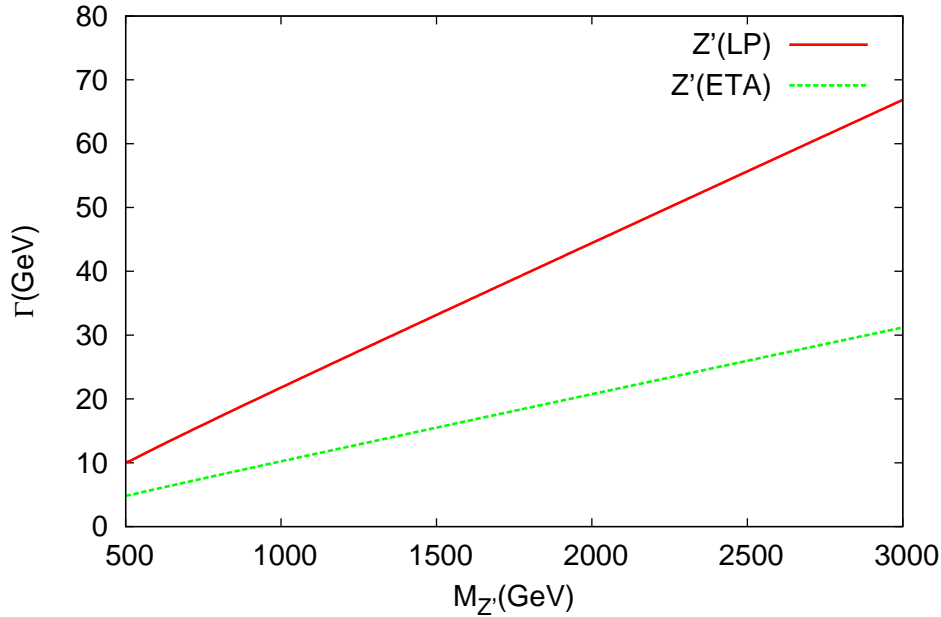}

\caption{Decay width of $Z'$ boson depending on its mass for two different
models. \label{fig:fig1}}
\end{figure}

\begin{figure}
\includegraphics{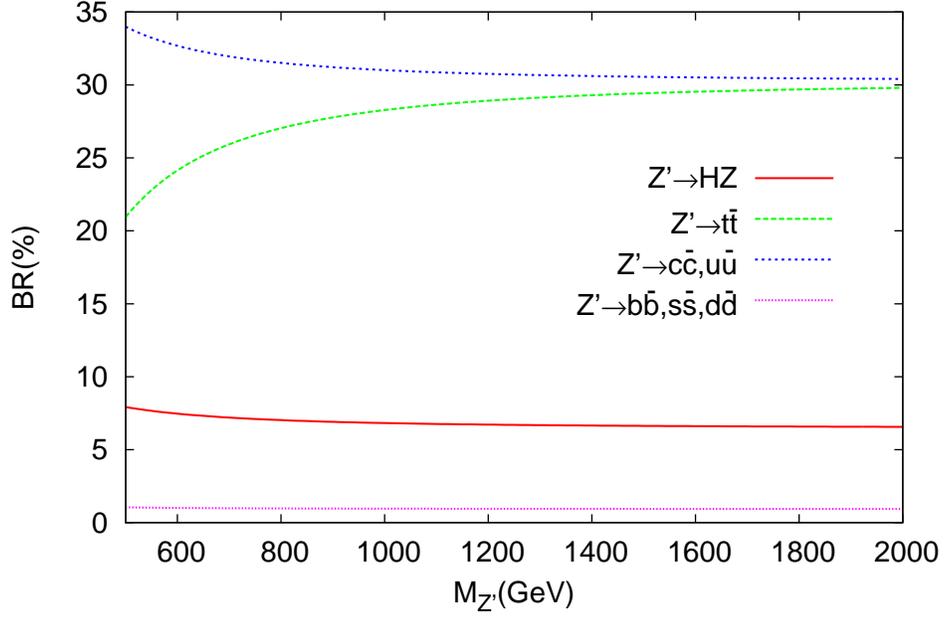}

\caption{Branching ratios of the $Z'$ boson as predicted by lepto-phobic model
(LP) depending on its mass. \label{fig:fig2}}
\end{figure}

\begin{figure}
\includegraphics{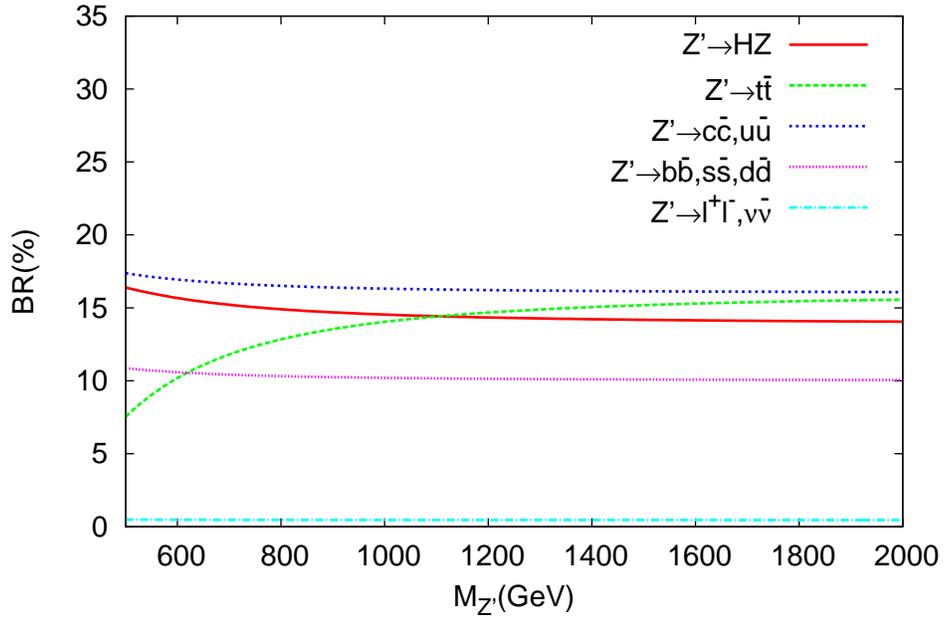}

\caption{Branching ratios of the $Z'$ boson as predicted by $\eta$ (ETA)
model of $E_{6}$ depending on its mass. \label{fig:fig3}}
\end{figure}

The production processes have contribution from the resonance production
of $Z'$ boson. Therefore, we already included the decay width of
the $Z'$ boson in the calculation within different $Z'$ models,
as shown in Fig. \ref{fig:fig1} for leptophobic (LP) and $\eta$
(ETA) models. The branching ratios versus $Z'$ mass are given in
Fig. \ref{fig:fig2} and Fig. \ref{fig:fig3} for the LP and ETA models,
respectively. 

\begin{figure}
\includegraphics{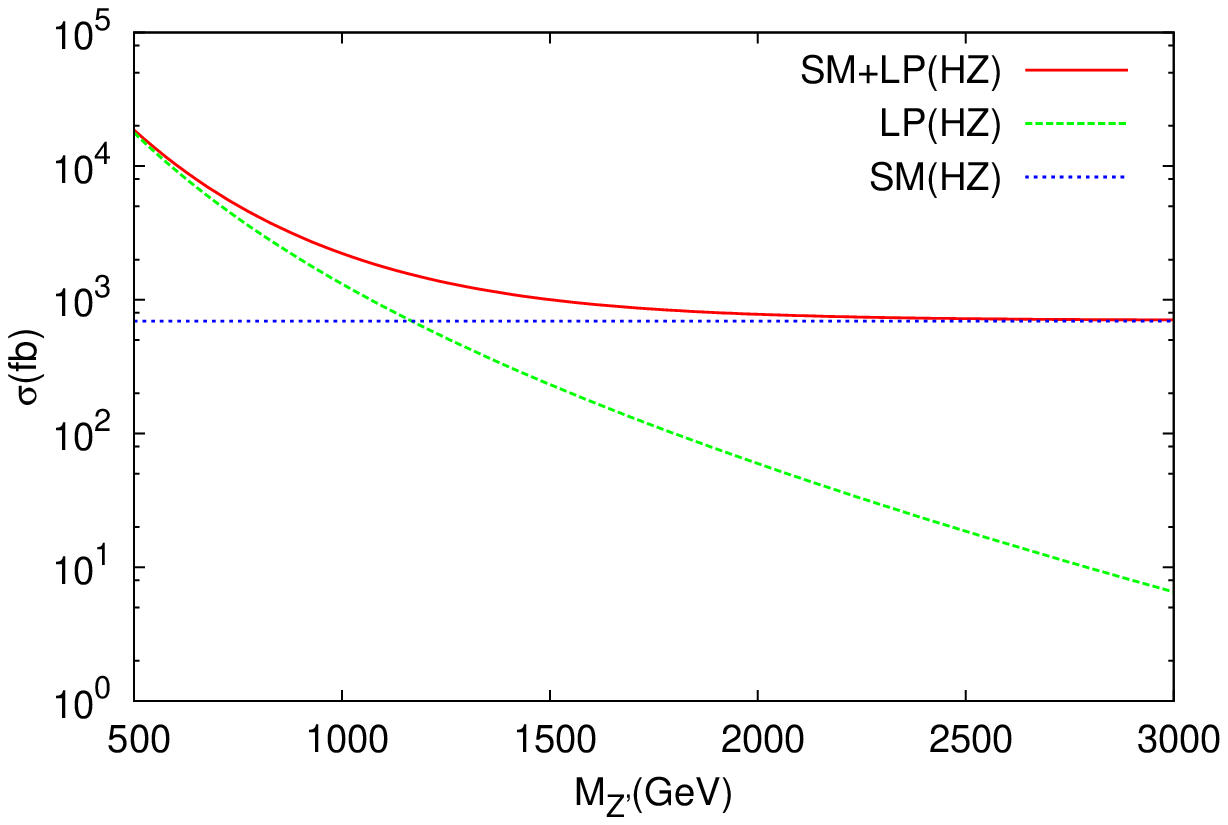}

\caption{The cross section for the process $pp\to HZX$ depending on the mass
of $Z'$ boson for LP model. \label{fig:fig4}}
\end{figure}

\begin{figure}
\includegraphics{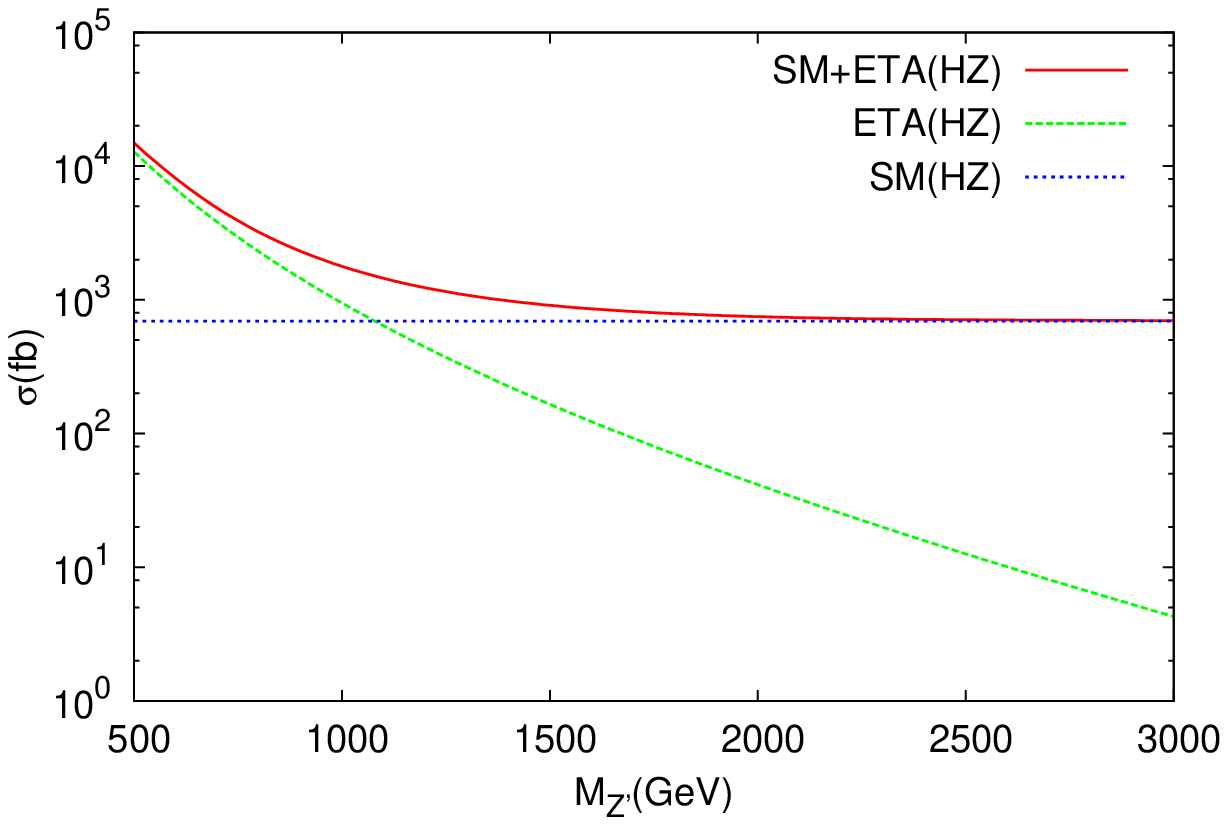}

\caption{The cross section for the process $pp\to HZX$ depending on the mass
of $Z'$ boson for ETA model. \label{fig:fig5}}
\end{figure}

\begin{figure}
\includegraphics{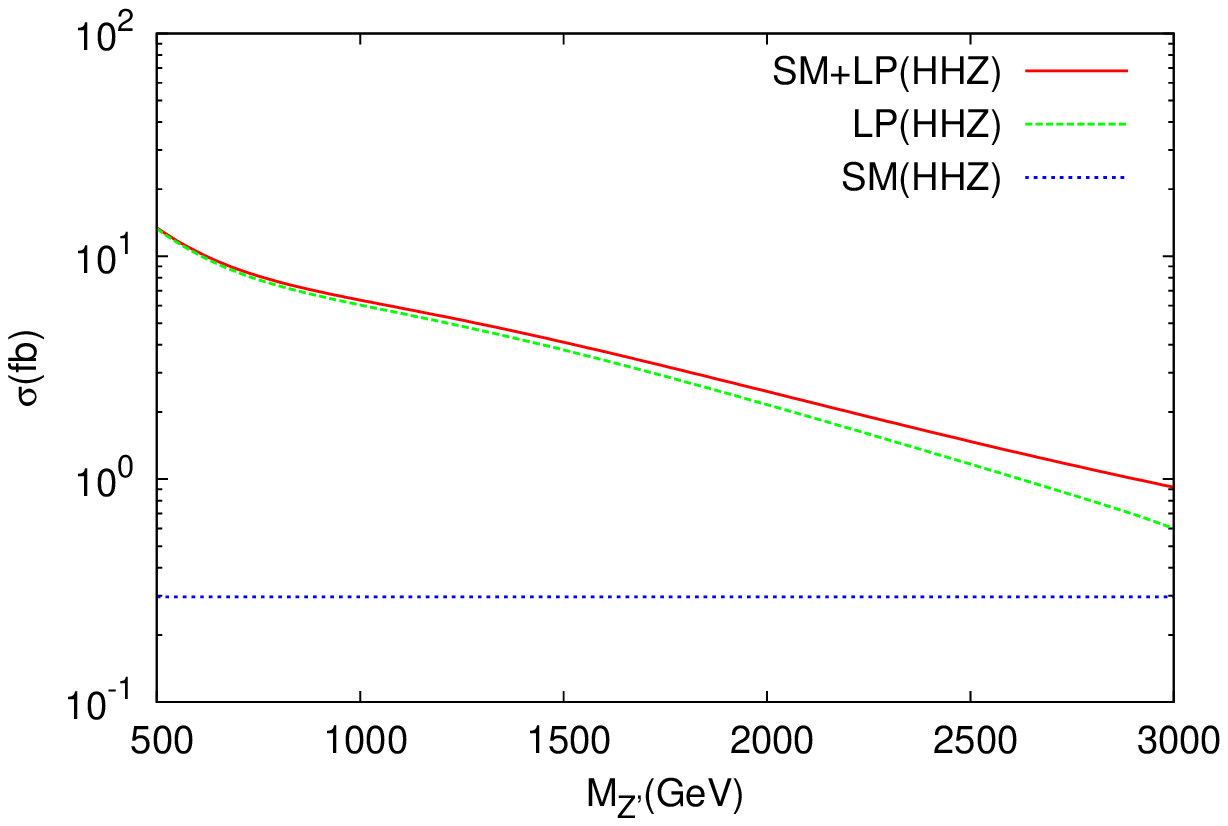}

\caption{The cross section for the process $pp\to HHZX$ depending on the mass
of $Z'$ boson for LP model. \label{fig:fig6}}
\end{figure}

\begin{figure}
\includegraphics{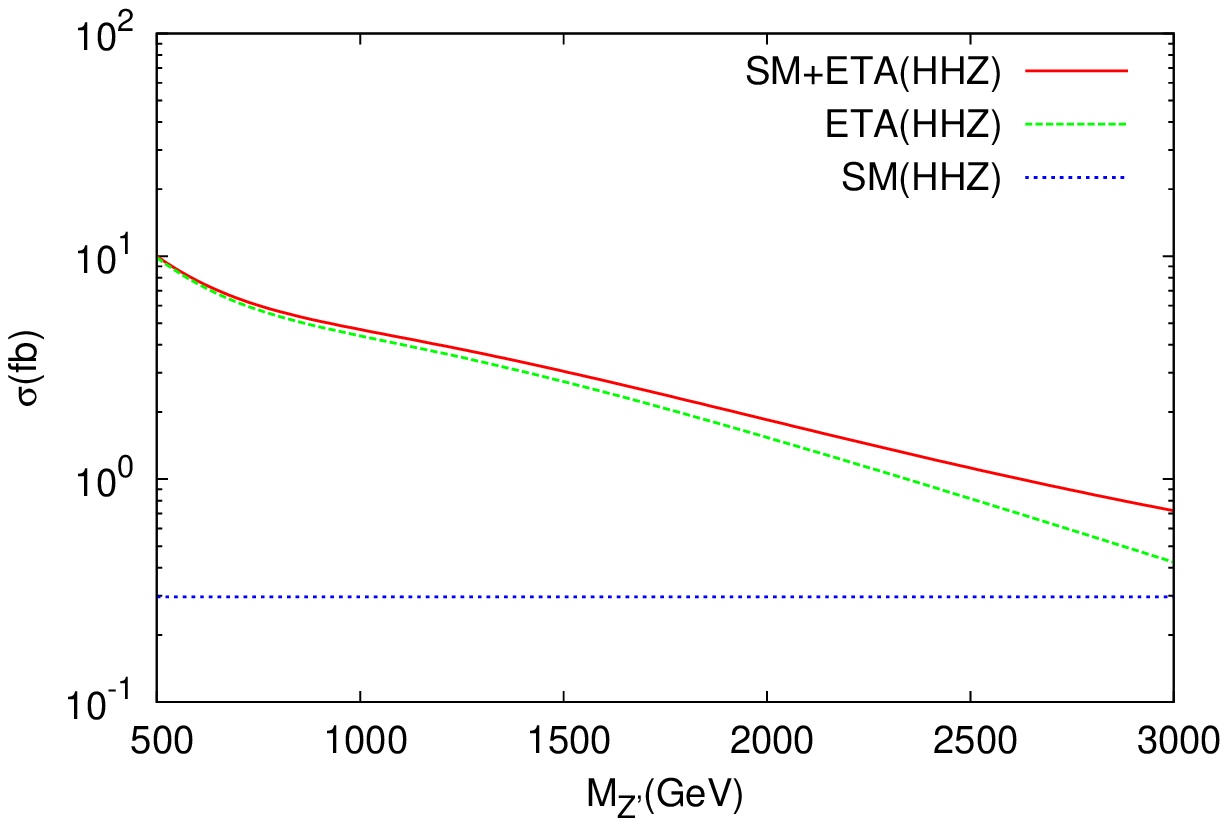}

\caption{The cross section for the process $pp\to HHZX$ depending on the mass
of $Z'$ boson for ETA model. \label{fig:fig7}}
\end{figure}

\begin{figure}
\includegraphics{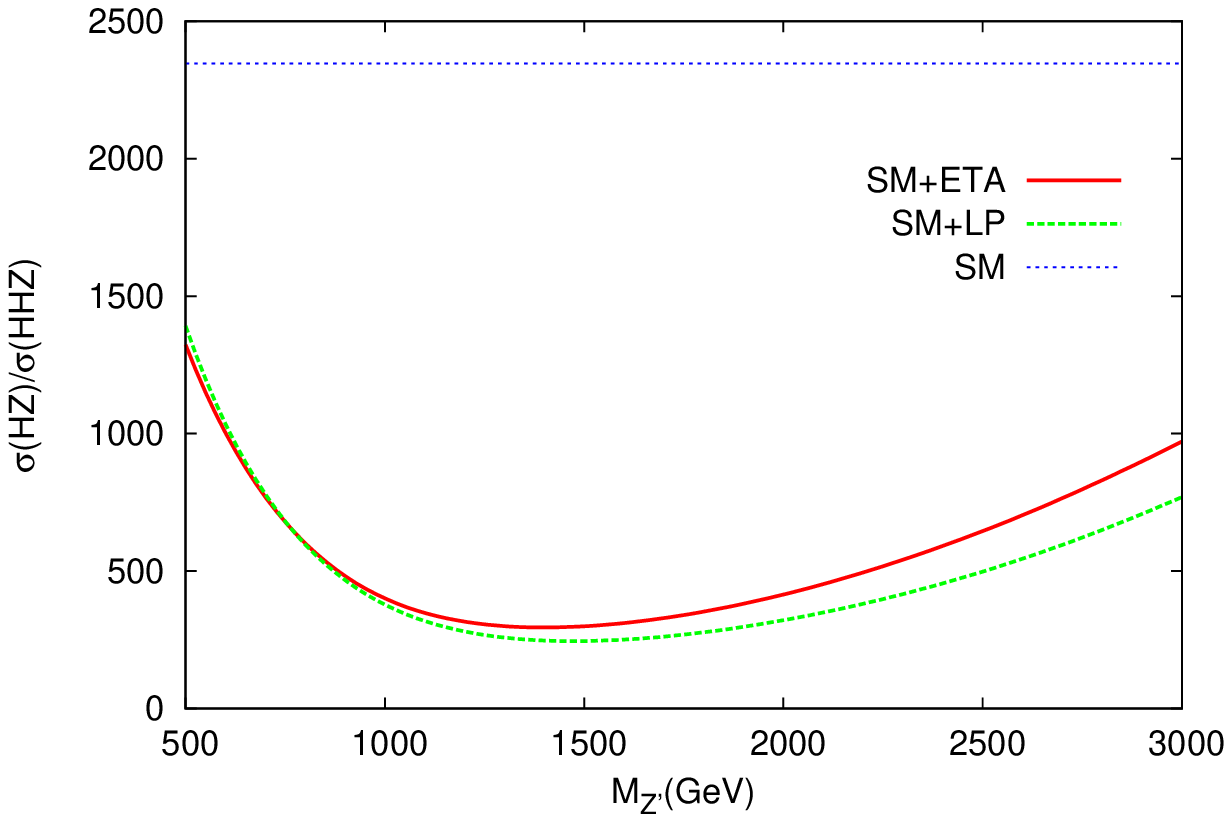}

\caption{The ratio of the cross sections for $HZ$ and $HHZ$ productions depending
on the mass of $Z'$ boson for two different $Z'$ models ETA and
LP. It is also shown the ratio of the SM cross sections. \label{fig:fig8}}
\end{figure}

At the LHC with $\sqrt{s}=14$ TeV, the cross section of the $HZ$
production through the signal process is about $0.20$ pb and $0.15$
pb with $Z'$ boson mass $m_{Z'}=1.5$ TeV for the LP and ETA model,
respectively. The cross section of $HHZ$ production is around $3.85$
fb and $2.78$ fb with $Z'$ boson mass $m_{Z'}=1.5$ TeV for the
LP and ETA model, respectively. As shown in Figs. \ref{fig:fig4}
- \ref{fig:fig5} and \ref{fig:fig6} - \ref{fig:fig7}, the background
cross sections result in $0.7$ pb and $0.29$ fb for $HZ$ and $HHZ$
production at $\sqrt{s}=14$ TeV, respectively. The ratios of the
cross sections for $pp\to HZX$ and $pp\to HHZX$ processes depending
on the mass of $Z'$ boson for two different $Z'$ models ETA and
LP are presented in Fig. \ref{fig:fig8}. This figure also includes
the ratio of the SM cross sections for given processes. The curves
shown in Fig. \ref{fig:fig8} represent large deviations from the
SM line depending on the $Z'$ boson mass. While the mass of $Z'$
boson is increasing the model difference becomes more apparent. The
final states of $2l+2b_{jet}$ and $2l+4b_{jet}$ are the main backgrounds
for the $HZ$ and $HHZ$ productions. In order to suppress the relevant
background, we can apply the invariant mass cuts, on dileptons to
be around the $Z$ boson mass, on dijets with each jets as \textbf{$b$}-tagged
to be around the Higgs mass.

\section{Parameter Space Analysis}

\begin{figure}
\includegraphics{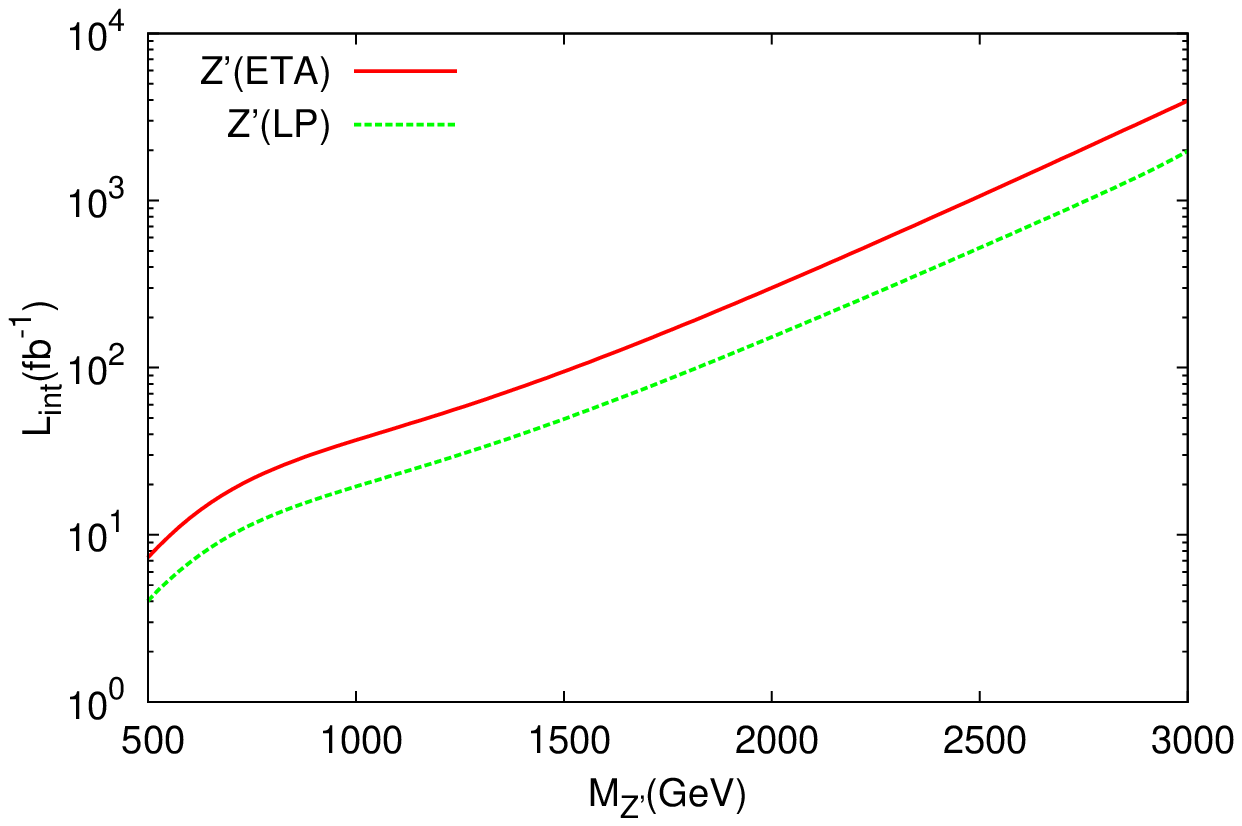}

\caption{Luminosity requirement for $3\sigma$ signal observability for final
state $2l+2b_{jet}$ depending on the mass of $Z'$ boson for two
different $Z'$ models ETA and LP. \label{fig:fig9}}
\end{figure}

\begin{figure}
\includegraphics{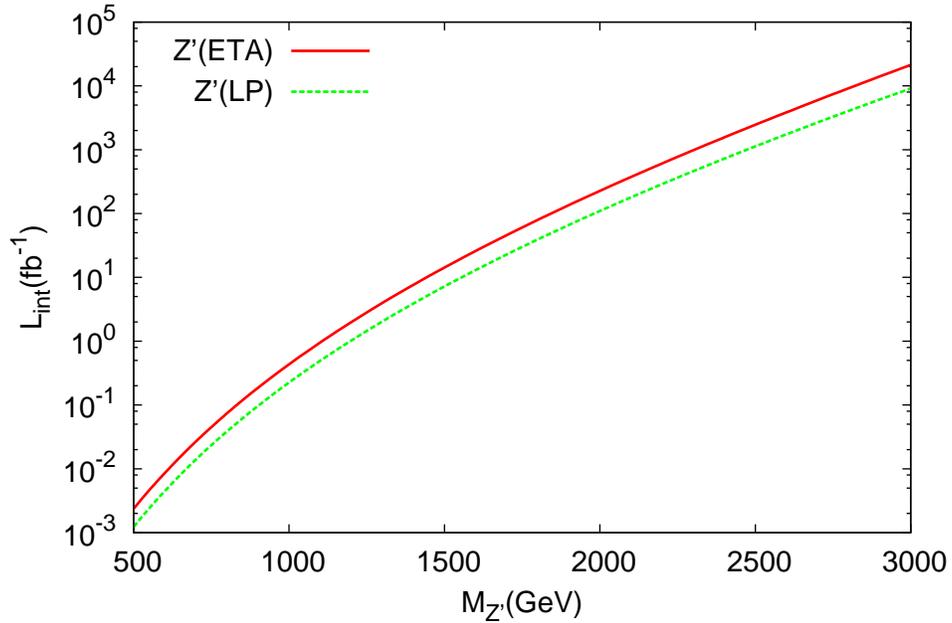}

\caption{Luminosity requirement for $3\sigma$ signal observability for final
state $2l+4b_{jet}$ depending on the mass of $Z'$ boson for two
different $Z'$ models ETA and LP. \label{fig:fig10}}
\end{figure}

In the analysis, we use the cross section of the signal and background
for the final states explained in the previous section. A pair of
$b$-jets give an invariant mass distributions peak around the Higgs
mass and the dilepton invariant mass distribution give peak around
the $Z$ mass. It is also helpful to use angular seperations of the
leptons, $b$-jets and between leptons and jets. The massive $Z'$
boson can be reconstructed from the resonance peak in the invariant
mass spectrum of the final states $2l+2b_{jet}$ and $2l+4b_{jet}$
originating from the $HZ$ and $HHZ$ productions, respectively. The
luminosity needs for $3\sigma$ signal observability for final state
$2l+2b_{jet}$ depending on the mass of $Z'$ boson for two different
$Z'$ models ETA and LP are given in Fig. \ref{fig:fig9}. Having
an integrated luminosity of $L_{int}=100$ fb$^{-1}$ at the LHC with
$\sqrt{s}=14$ TeV, it is possible to search for $Z'$ boson up to
a mass value of $M_{Z'}=1.9$ TeV and $2.0$ TeV for the ETA and LP
models as seen from Fig. \ref{fig:fig9}. The search range for the
$Z'$boson mass can be found as $M_{Z'}=1.5$ TeV and $1.8$ TeV for
ETA and LP models for $2l+4b_{jet}$ as given in Fig. \ref{fig:fig10}.

\section{Conclus\i{}ons}

We present the attainable parameter space for two different $U(1)'$
models by studying the processes $pp\to HZX$ and $pp\to HHZX$ at
the LHC with $\sqrt{s}=14$ TeV. The $HZ$ production has large cross
section and it has the advantage of two \textbf{$b$}-jets and two
charged leptons in the final state. The background for $HHZ$ production
has lower cross section than $HZ$ production, however it has the
advantage of four \textbf{$b$}-jets and two charged leptons in the
final state. The Higgs boson mostly decays into two $b$-jet channel
and we use the advantage of efficient identification of dileptons
from $Z$ boson. The couplings of $Z'$ boson to the SM quarks play
a role in the production, and the $U(1)'$ charge of the Higgs boson
scales the $HZZ'$ coupling. The searches for $HVV$ couplings will
also extend our perspectives for new physics at the LHC with $\sqrt{s}=14$
TeV and $L_{int}=100$ fb$^{-1}$.

\end{document}